\documentclass[amsmath, amssymb,pra,preprint,aps]{revtex4-2}
\usepackage{graphicx,float}
\usepackage{color}
\usepackage{mathrsfs}
\usepackage{hyperref}
\usepackage{lineno}
\usepackage{tikz}
\usepackage{url}
\tikzset{every picture/.style={remember picture}}
\usepackage{colortbl}
\usepackage{ulem}
\usepackage{xcolor} \usepackage{ulem} \usepackage{changebar}

\colorlet{ins}{blue} \colorlet{del}{red}

\begin{document}
	
	\title{Fast dispersion tailoring of multi-mode photonic crystal resonators}
	\author{Francesco Rinaldo Talenti$^{1,2,*}$, Stefan Wabnitz$^{1,3}$, In\`es Ghorbel$^{2}$, Sylvain Combri\'{e}$^{2}$, Luca Aimone-Giggio$^2$, Alfredo De Rossi$^{2}$}
	\affiliation{$^{1}$Dipartimento di Ingegneria dell’Informazione, Elettronica e Telecomunicazioni, Sapienza University of Rome, 00184 Rome, Italy.\\
		$^{2}$ Thales Research and Technology, Campus Polytechnique, 1 Avenue Augustin Fresnel, 91767 Palaiseau, France.\\
		$^{3}$ CNR-INO, Istituto Nazionale di Ottica, Via Campi Flegrei 34, 80078 Pozzuoli (NA), Italy.
		$^{*}${Corresponding author: francescorinaldo.talenti@uniroma1.it}}

	\begin{abstract}
		We introduce a numerical procedure which permits to drastically accelerate the design of multimode photonic crystal resonators. Specifically, we demonstrate that the optical response of an important class of such nanoscale structures is reproduced accurately by a simple, one-dimensional model, within the entire spectral range of interest. This model can describe a variety of tapered photonic crystal structures. Orders of magnitude faster to solve, our approach can be used to optimize certain properties of the nanoscale cavity. Here we consider the case of a nanobeam cavity, where the confinement results from the modulation of its width. The profile of the width is optimized, in order to flatten the resonator dispersion profile (so that all modes are equally spaced in frequency). This result is particularly relevant for miniaturizing parametric generators of non-classical light, optical nano-combs and mode-locked laser sources. Our method can be easily extended to complex geometries, described by multiple parameters.
	\end{abstract}    
	
	\maketitle
	
	\section{Introduction}
	The nonlinear interaction among several resonant fields in an optical resonator leads to efficient Raman and Brillouin scattering, three and four-wave-mixing, optical parametric oscillation\cite{boyd2020}, laser mode locking and frequency comb generation\cite{kippenberg2018}. Scaling down the size of optical resonators implies that the optical power level for triggering nonlinear effects decreases as $V^{-1}$ or $V^{-2}$, where $V$ is an effective volume of the spatial distribution of the interacting fields. In the context of photonic integration, the decrease of the power budget is of paramount importance.\\
	Nanoscale optical resonators such as photonic crystals are able to confine light within $V\approx \lambda^3$, i.e., a wavelength-sized volume, with a photon decay time, or interaction time, well above 1 ns (i.e., the cavity quality factor $Q \gg 10^6$). Owing to these properties, it has been possible to demonstrate nanoscale lasers\cite{matsuo2010, Crosnier2017, yu2017,nozaki2019}, Raman sources\cite{takahashi2013} and, more recently, optical parametric oscillators\cite{marty2021}, all operating with a power supply (optical or electrical) in the $\mu$W range. Yet, a major challenge remains in achieving the nonlinear interaction of multiple longitudinal modes, as it occurs in mode-locked lasers or in micro-combs. While ring or microdisk resonators naturally provide the necessary, nearly frequency equispaced set of cavity resonances, achieving the same condition in nanoscale resonators is notoriously a nontrivial task. On the other hand, nanoscale resonators could, in principle, be designed in a way that a specified number of modes, starting from the fundamental, and only these, are allowed to take part to a nonlinear interaction. This unique property implies not only that a much higher degree of control on power transfer among modes (which is crucial in quantum and signal processing applications\cite{stone2022}) can be achieved, but also leads to maximizing the interaction efficiency. This is because, in the typical configuration of a nanoscale resonator, the lowest order modes are also the most tightly confined. Moreover, in a mode-locked nanolaser, the control of the interacting modes would enable a favorable scaling of repetition rate vs. the size of the device\cite{sun2019}.\\
	It has been shown that some specific designs of a photonic crystal cavity lead, for some set of parameters, to frequency equispaced eigenmodes; moreover, their mode envelopes are described by Hermite-Gauss functions. This suggests that, within a certain spectral range, the complex photonic crystal structure can be well approximated by a quantum-mechanical harmonic oscillator model\cite{combrie2017,marty2019}. It has also been shown that post-fabrication trimming is effective in correcting for fabrication tolerances, thereby demonstrating an almost perfect alignment of the cavity resonances\cite{clementi2019}. Yet, a systematic design approach for generating a given number of equispaced modes, or, more generally, with a prescribed dispersion profile, while at the same time maximizing the radiation-limited Q-factor, is still missing, while brute-force methods are extremely inefficient.\\  
	Finding a cavity geometry, or more generally, a physical system whose response to an input excitation corresponds to a well-defined target function, e.g., a spatial distribution of the dielectric permittivity such that the electromagnetic field has prescribed resonances, belongs to the class of \textit{inverse problems}, which are notoriously difficult to solve. Yet, the progress of nanofabrication techniques has motivated the development of powerful methods such as topological optimization (TO) \cite{jensen2011} and inverse design (ID) \cite{molesky2018}. The common feature of these two approaches is that their result is a spatial distribution of $\varepsilon(\mathbf{x})$, rather than an optimized set of parameters for a pre-defined geometry. These methods are therefore able to \textit{create} novel geometries, hence the reference to \textit{design}. Moreover, automatic differentiation\cite{minkov2020} and the adjoint method\cite{hughes2018} enable a very efficient computation of the gradient, which is required in the iterative search of the optimum distribution, even in the presence of nonlinearity.\\
	%FIGURE 1
	\begin{figure}[h!]
		\begin{minipage}[c]{0.5\textwidth}
			\includegraphics[width=\textwidth]{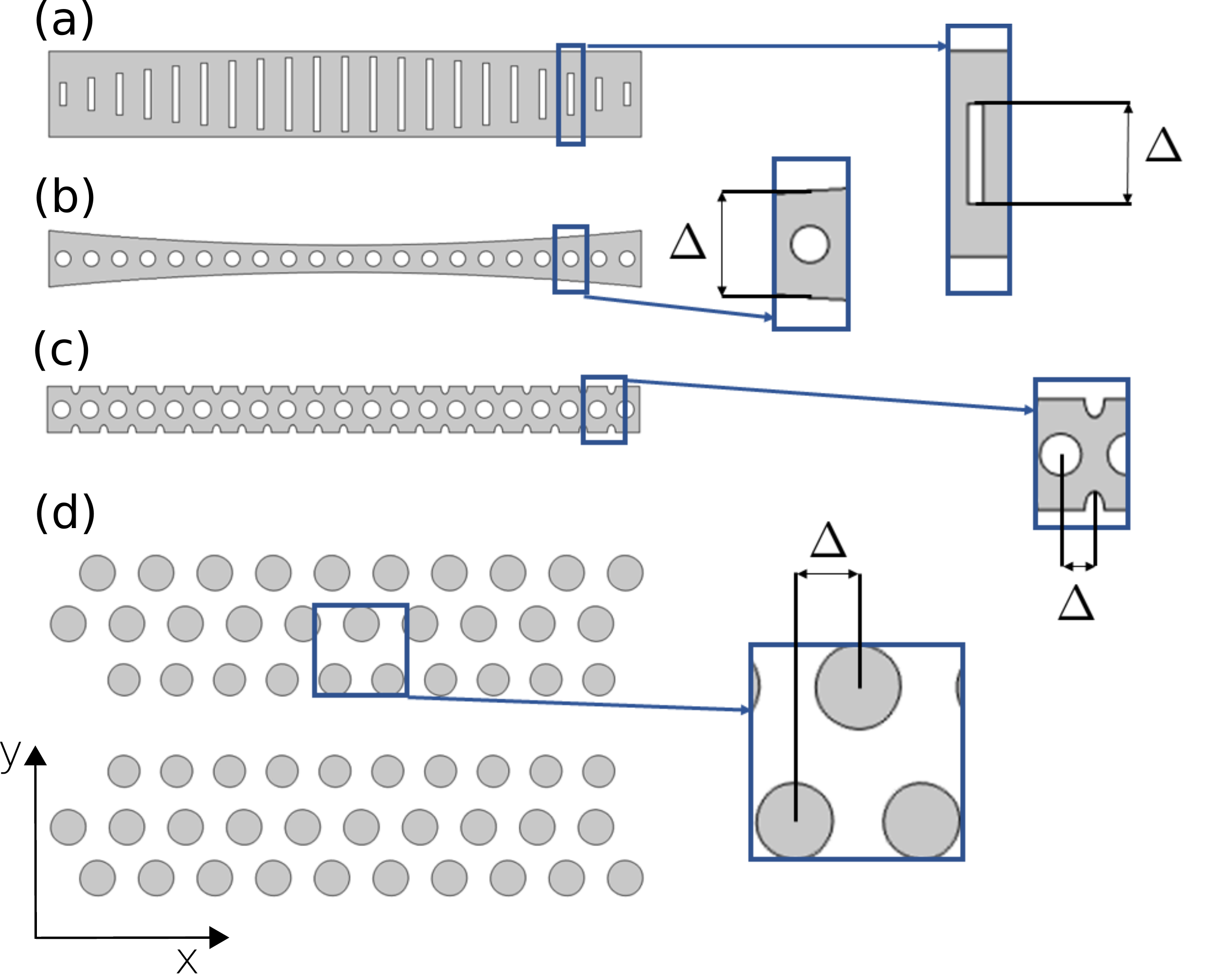}
		\end{minipage}\hfill
		\begin{minipage}[c]{0.48\textwidth}
			\caption{Common design of high-Q resonators based on \textit{gentle} confinement: (a) tapered Distributed Feedback Grating\cite{santis2014}, (b) 1-D nanobeam with parabolic width\cite{ByeongAhn2010}, (c) 1-D nanobeam bichromatic\cite{ghorbel2019}, and (d) 2-D bichromatic resonator\cite{combrie2017}, and corresponding tapering parameter $\Delta$. }
			
			\label{fig:geometry}
		\end{minipage}
	\end{figure}

	Here we follow a radically different approach, which is arguably more suited for the class of problems under consideration. This is motivated by the fact that the geometries of nanoscale resonators with the largest experimentally reported Q factors\cite{asano2017,notomi2007,quan2011,santis2014,bazin2014} are still based on the principle of \textit{gentle confinement}\cite{Akahane2003}. In other words, these nanostructures are essentially periodic, with an adiabatic tapering of some parameters, i.e., a gentle change of the radius of the holes, the period or the magnitude of a "dislocation" defect, etc.. We note that more aggressive design strategies, including TO or ID, have instead been considered for different tasks, e.g., for maximizing light-matter interactions in \textit{single-mode} resonators\cite{minkov2020,Wang2018}. \\
	Let us restrict our search to a family of structures which can be described by means of a periodic pattern $\varepsilon(\mathbf{x},\Delta)$ that depends on a \textit{control} parameter $\Delta$, which is supposed to  adiabatically vary in space (i.e., \textit{gently}). Some examples of such geometries are given in Fig. \ref{fig:geometry}. The crucial point is that it is possible to map the three-dimensional (3D) Maxwell equations (ME) into an equivalent system of one-dimensional (1D) equations, which will be referred to as the \textit{reduced model} (RM). Remarkably, the relative precision of the resonances predicted by the RM turns out to be at least as good as the precision of the direct numerical solution of the 3D ME. The search of the desired optimal spatial dependence of $\Delta$ will be performed by using any suitable optimization method, leveraging on the extremely faster solution of the RM, when compared the direct solution of the 3D ME. The RM in itself only requires a single direct solution of the 3D ME, for building an initial approximation of the structure. Subsequent applications of the RM are used, in order to refine the first approximation. As we shall see, in total only three 3D solves are sufficient for obtaining a design that matches our target, with an accuracy that is equivalent to that of directly solving the 3D ME, but with a comparatively much larger number of iterations.\\
	Hereafter we will first discuss the derivation of the RM, then we will formulate a design target followed by the introduction of the optimization procedure, including model calibration. Finally, we will discuss possible applications and generalizations of the model.
	\section{Reduced model for a periodic photonic crystal}
	%
	%FIGURE 2
	\begin{figure}[h!]
		\begin{minipage}[c]{0.5\textwidth}
			\includegraphics[width=\textwidth]{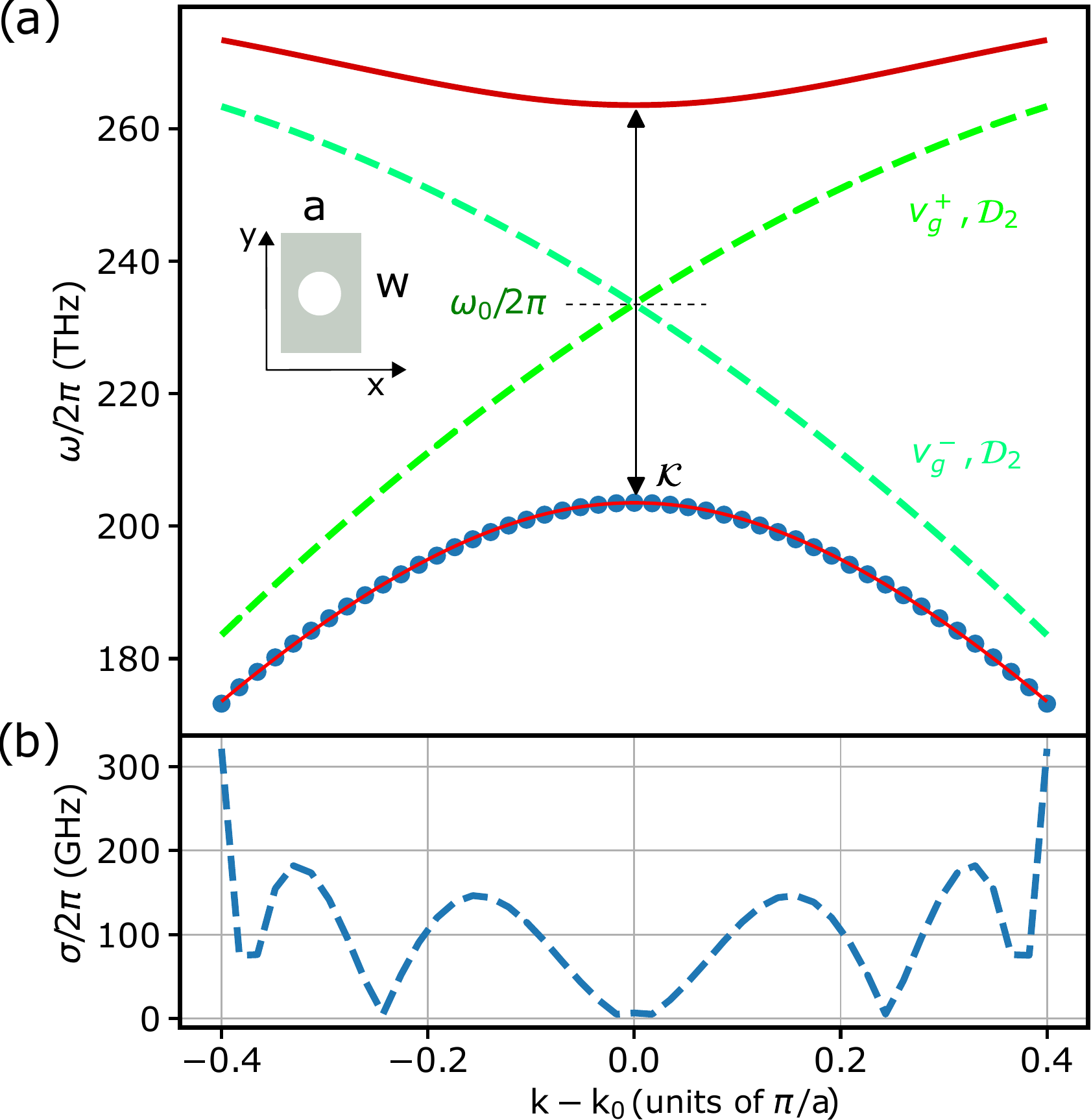}
		\end{minipage}\hfill
		\begin{minipage}[c]{0.48\textwidth}
		\caption{(a) Dispersion diagram of a periodic structure (see inset) with $w=450$nm and period $a=465$nm, centered in the $K$ point of the reduced Brillouin zone ($k_0=\pi/a$). The reduced model considers coupled forward and backward waves (dashed lines), generating the valence and conduction bands (solid lines). The filled circles represent the valence band calculated by periodic 3D ME. (b) The corresponding residuals $\sigma/2\pi$ of the fit.}
	\label{fig:reduced_model}
		\end{minipage}
	\end{figure}

	The reduced model is inspired by the so-called $\mathbf{k} \cdot  \mathbf{p}$ method \cite{YuCadorna2005}, which is used in solid state physics to model the electronic band structure of crystals. The main idea of the method is to describe the dispersion relation of the electronic bands (i.e., electron energy vs. wavevector $\mathbf{k}$) through a suitable algebraic equation, which is built upon the eigenfunctions of the exact Hamiltonian at the bands extrema (at points of high symmetry, e.g., $\mathbf{k}=0$, or gamma point). Within a range of energies of interest, the dispersion relation is extrapolated from the gamma point by treating the $\mathbf{k}\cdot \mathbf{p}$ term as a perturbation. In this way, the complexity of solving the full Schr\"{o}dinger equation for the crystal is reduced by using a much simpler model, where only a few parameters need to be suitably adjusted. As a consequence of this approach, a local modulation of a semiconductor, e.g., of a heterostructure, can be well described in terms of a change of these parameters within the energy range of interest. As a result, it is possible to introduce a much simpler Schr\"{o}dinger equation, which only depends on these parameters.\\
	In optics, the simplest model describing the propagation of waves in a periodic dielectric is provided by the case of a distributed Bragg reflector. Here, a modulation with period $\Lambda$ of the dielectric permittivity couples  forward and backward waves. A simple algebraic equation approximates the dispersion in the spectral range that is centered at the Bragg angular frequency $\omega_0=\pi c n^{-1}\Lambda^{-1}$. Here $c$ is the speed of light and $n$ is an effective refractive index,  which describes the optical field distribution as a result of the dielectric inhomogeneity \cite{YarivQE}. In the presence of an intensity dependent contribution to the refractive index, one obtains coupled wave propagation equations which generalize the Massive Thirring model of field theory; their solitary wave solutions (gap or Bragg solitons) describe the localisation of wavepackets in periodic media\cite{AcevesWabnitz1989}. The simplicity of the gap soliton model has facilitated the study of soliton stability by using analytical criteria\cite{derossi1998}.  
	% the thirring model describes self-interactions in  Dirac field https://en.wikipedia.org/wiki/Quantum_field_theory
	%https://en.wikipedia.org/wiki/Thirring_model
	% massive thirring
	%\begin{equation}
	%L = \int\limits dx \left [ i {\overline \Psi \; } \gamma ^ \mu \partial _ \mu \Psi - m _ {0} {\overline \Psi \; } \Psi - { %\frac{g}{2} } : [ {\overline \Psi \; } \gamma _ \mu \Psi ] ^ {2} : \right ] , 
	%\end{equation}
	Soliton dynamics has been experimentally demonstrated in nanoscale photonics\cite{colman2010}, and it has been shown that nonlinear coupled waves models are able to fully capture the underlying physics\cite{malaguti2012}. Hereafter, we will only consider a generalized linear version of the gap soliton model, and demonstrate that the model accurately describes wave propagation in adiabatically modulated photonic crystal structures.\\ 
	%
	%       coupled modes
	%
	Let us consider two counter-propagating waves $E^{\pm}(x,t)=A_k^{\pm}\exp(\imath\omega t \pm \imath k x)$, with group velocity $v_g$, and coupled by a periodic modulation of the dielectric permittivity with scaled magnitude $\mathcal{K}$. In the presence of this linear coupling, the dispersion relation of the waves is described by the coupled equations:
	\begin{eqnarray}
	%    \begin{cases}
	\nonumber
	(v_g k+\mathcal{D}_2k^2+\omega_0)A_k^+ + \mathcal{K}A_k^-=\omega A_k^+\\
	(-v_g k+\mathcal{D}_2k^2+\omega_0)A_k^- + \mathcal{K}A_k^+=\omega A_k^-    
	%    \end{cases},
	\label{eq:reduced_model}
	\end{eqnarray}
	Let us note that we introduced the Bragg angular frequency $\omega_0$, and added the second-order dispersion term $\mathcal{D}_2$. The set  $\{\omega_0, \mathcal{K}, v_g, \mathcal{D}_2\}$ describes the dispersion of the coupled waves, and we will refer to it as the \textit{structure parameters}. We now derive these parameters for periodic structures with different widths $w$, by focusing on a specific portion of the dispersion relation, namely one or more bands, as shown in Fig. \ref{fig:reduced_model}(a). 
	
	Here we consider a so-called nanobeam photonic crystal, where the width $w$ of the beam takes the role of the control parameter (Fig. \ref{fig:geometry}(b)). The nanobeam is supposed to be made out of a III-V group semiconductor alloy In$_{0.5}$Ga$_{0.5}$P, with refractive index $n=3.17$. The nanobeam is $h=180$ nm thick, with a $w=450$ nm width, and a $a=465$ nm period; the holes radius is $0.27a$. The valence band $\omega_v(k)$ is obtained by solving the 3D ME with Bloch periodic boundary conditions along $x$, i.e., $\mathbf{E}_k(\mathbf{r}+a\hat{x})=E_k(\mathbf{r})\exp{(\imath k a)}$. This is performed by means of the periodic FDTD (finite difference in time domain) algorithm with a perfectly matched layer placed at the $z$ and $y$ boundaries. The parameters are adjusted in order to minimize the error $N^{-1}\sum_{i=0}^{N-1}{\sigma^2(k_i)}$ over the $N$ points in the reciprocal space, with $\sigma^2(k_i)=[\omega_v(k_i) - \omega_v^{(RM)}(k_i)]^2$ obtained from the reduced model $\omega_v^{(RM)}$, i.e. the characteristic equation solutions of the linear system eq. \ref{eq:reduced_model}. This generates the set of parameters $\{\omega_0, \mathcal{K}, v_g, \mathcal{D}_2\}(w)$, which depends on the control parameter $w$. The average of the residual error in the reciprocal space region of interest $\sigma_{fit}=\overline{\sigma(k)}$ is about $100$ GHz, cfr. Fig. \ref{fig:reduced_model}(b). Let us note that this error is about the same as the estimated discretization error of the FDTD method\cite{deLasson2018}. This point is further discussed in the Appendix.\\
	Fig. \ref{fig:RM_vs_parameter} describes the dependence of the structure parameters on the control parameter $w$. Panel (a) shows that the fit error decreases when $w$ grows from $0.45\mu$m to $0.52\mu$m, meaning that the dispersion relation is increasingly closer to that of the RM. In panels (c-f), the blue dashed line represents the polynomial fit of the extracted parameters with respect to $w$. Here, we make a crucial assumption, namely that the dependence of the structure parameters on $w$ is smooth. The figure shows that a low-order ($3^{rd}$) polynomial is a good approximation, moreover the residual of $\omega_0$ (blue solid line in panel b) is about  100 GHz or below. The result here is a set of polynomial coefficients $\mathcal{C}_i^{(\mathcal{P})}$ for each parameter $\mathcal{P}=\{\omega_0, \mathcal{K}, v_g, \mathcal{D}_2\}$. This two-step interpolation of the dispersion of the periodic structure removes the minute deviations which might be related to the discretization error.

	%FIGURE 2
		\begin{figure}[h!]
		\begin{minipage}[c]{0.5\textwidth}
			\includegraphics[width=\textwidth]{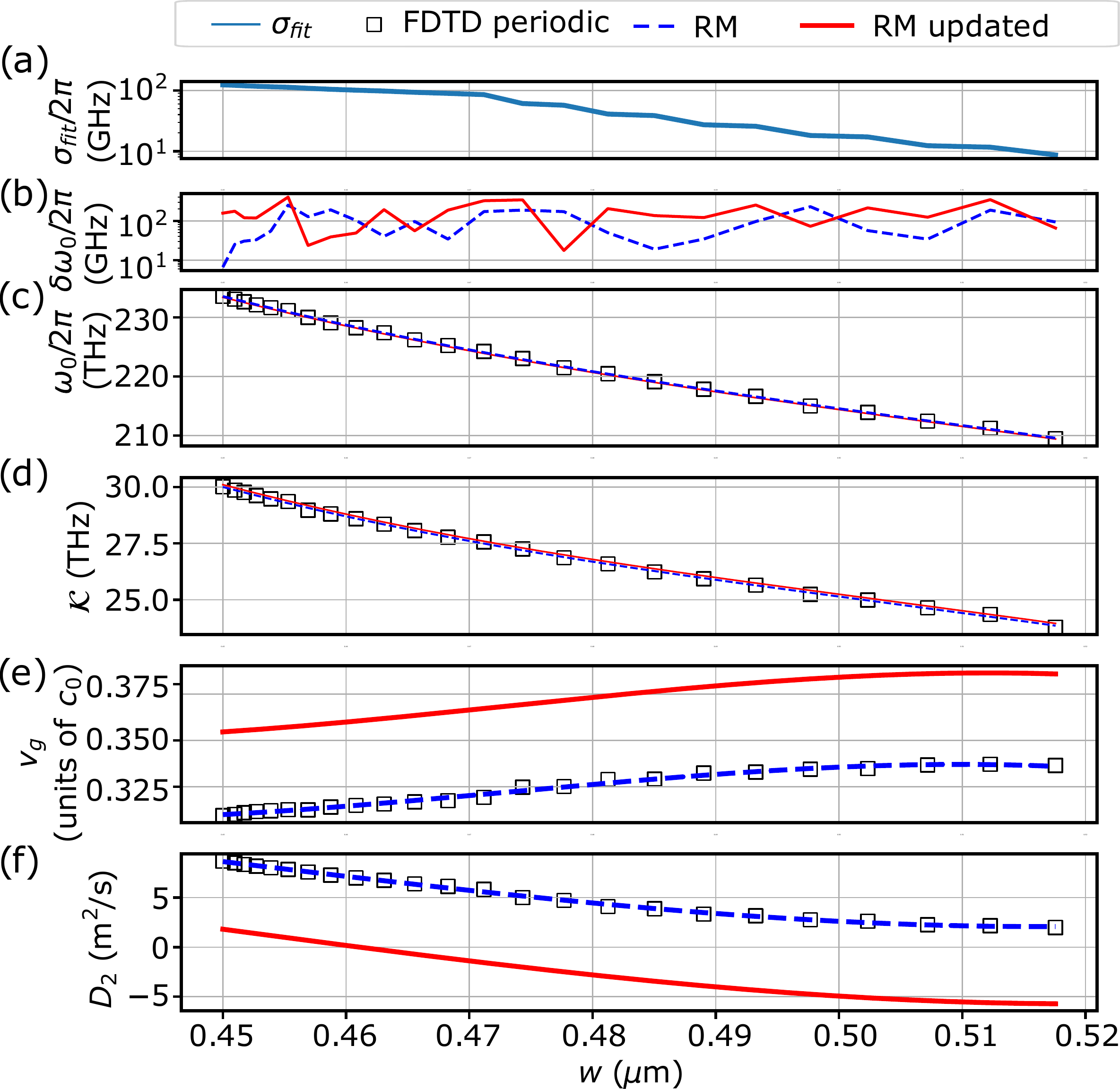}
		\end{minipage}\hfill
		\begin{minipage}[c]{0.48\textwidth}
		\caption{Calculated structural parameters for the RM, vs. the control parameter ${w}$. (a) averaged residual of the fit; (c-f) extracted parameters (black squares), polynomial fit (blue dashed line) and polynomial fit on the updated RM (solid red). In (e) the group velocity is reported in units of $c_0$, i.e. the speed of light in vacuum. (b) residual of the fit of $\omega_0$ (dashed blue), and after updating the RM (red solid). 
}	
\label{fig:RM_vs_parameter}
		\end{minipage}
	\end{figure}

	\section{Reduced model of a tapered nanobeam cavity}
\begin{figure}[h!]
		\begin{minipage}[c]{0.5\textwidth}
			\includegraphics[width=\textwidth]{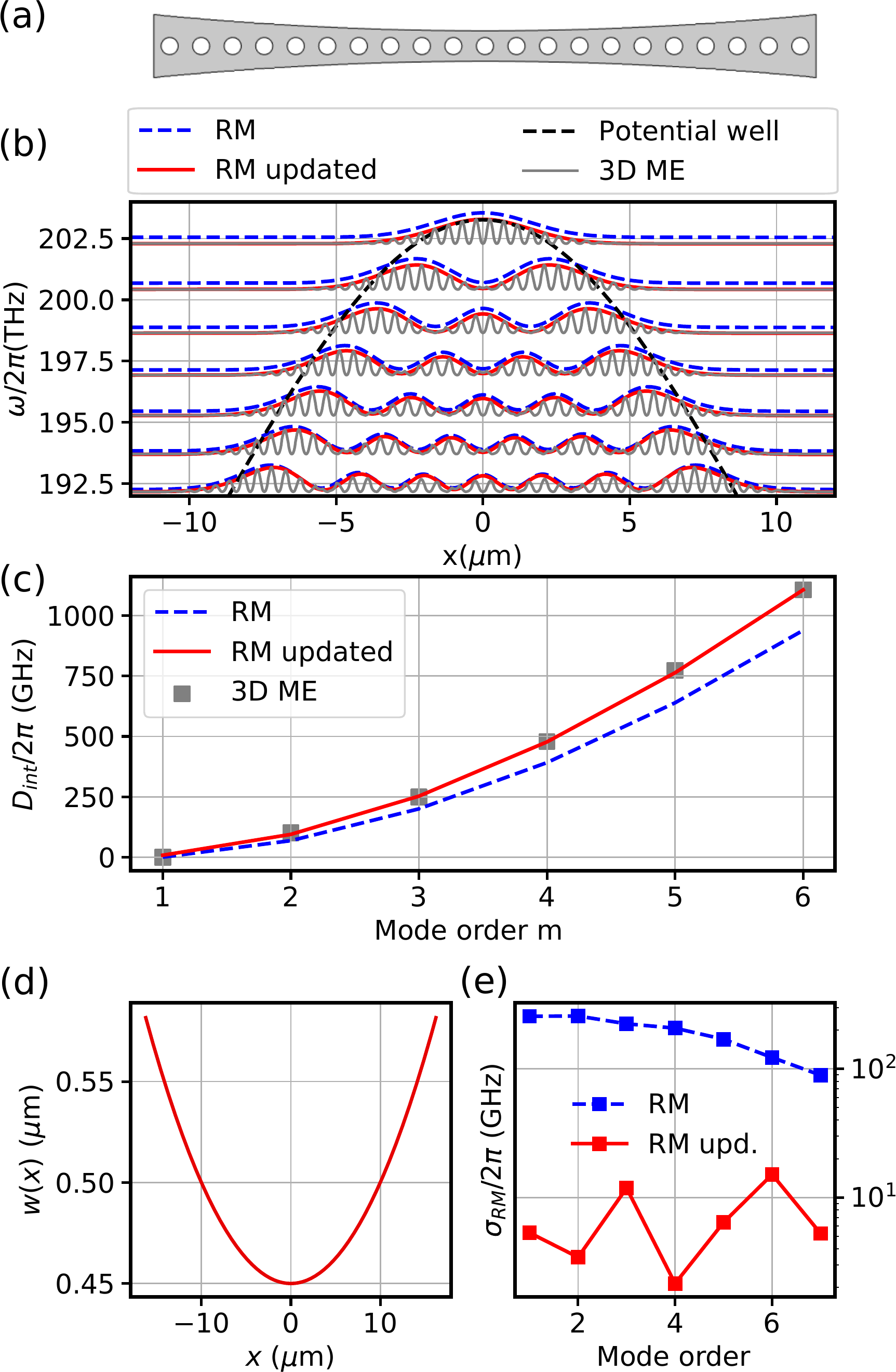}
		\end{minipage}\hfill
		\begin{minipage}[c]{0.48\textwidth}
		\caption{Tapered nanobeam optical cavity. (a)  simplified layout; (b) spatial dependence of the valence band edge at $k=k_0$ (dashed black line), normalized distribution of the squared electric field corresponding to the calculated Bloch modes (3D ME) of the cavity along the axis $y=0$, $z(d)=0$ (solid grey), envelopes calculated from the reduced model of the cavity (blue dashed) and after model update (red solid). The vertical offset corresponds to the frequency at resonance. (c) Integrated dispersion $D_{int}$ calculated by solving the 3D ME (squares), with the reduced cavity model (blue dashed line), and after update (red solid). (d) Spatial dependence of $w$. (e) Frequency deviation for each mode $\sigma_{RM}/2\pi$ of the RM (blue dashed) and the updated RM (red solid), relative to the 3D ME calculations)   }
\label{fig:nanobeam_cavity}
		\end{minipage}
	\end{figure}

	Let us now consider an optical resonator where the confinement is due to the tapering of the width $w(x)=\rho_0 + \rho x^2$ ($\rho>0$) of the nanobeam, as described in ref. \cite{ByeongAhn2010}. When considering the $w$ dependence of $\omega_0$ and $\mathcal{K}$ in Fig. \ref{fig:RM_vs_parameter}(d), it is immediate to realize that the edge of the valence band $\omega_{vb}=\omega_0-\mathcal{K}$ decreases as $w$ increases. This leads to localisation of Bloch waves in the valence band of the nanobeam, if $w$ is smaller in the center of the nanobeam.  Let us consider the case with $\rho=500$m$^{-1}$. Fig. \ref{fig:nanobeam_cavity}(b) shows the corresponding spatial distribution of the modes, as it is obtained from the solution of the 3D ME using the Finite Element Method (details are discussed in the Appendix). \\
	Let us now build a RM for the cavity, and define the linear operator $\mathcal{L}_{RM}$ acting on complex-valued functions of space $x$ ($\mathcal{R}\rightarrow\mathcal{R}^2$):
	\begin{equation}
	\mathcal{L}_{RM}=
	\begin{bmatrix}
	-\mathcal{D}_2\partial_x^2 + \imath v_g\partial_x  + \omega_0 &      \mathcal{K} \\
	\mathcal{K}     & -\mathcal{D}_2\partial_x^2 - \imath v_g\partial_x + \omega_0
	\end{bmatrix} ,
	\label{eq:RM_operator}
	\end{equation}
	Here, the structure parameters $\{\omega_0, \mathcal{K}, v_g, \mathcal{D}_2\}$ are all functions of $x$ via the profile $w(x)$ and the polynomials $\mathcal{C}_i^{(\mathcal{P})}$. Namely, for each parameter $\mathcal{P}$ the corresponding function of $x$ reads as
	\begin{equation}
	\mathcal{P}(x) = \sum_l \mathcal{C}_l^{(\mathcal{P})}w(x)^l
	\label{eq:RM_cavity_par}
	\end{equation}
	The polynomial expansion is replaced by constant values for $|x|>x_{max}$, namely $\mathcal{P}|_{|x|>x_{max}} =\mathcal{P}(x_{max})$. The eigenfunctions $\psi=[\mathcal{A}^+ , \mathcal{A}^-]$ of the equation:
	\begin{equation}
	(\mathcal{L}_{RM} - \omega)\psi=0
	\label{eq:RM_cavity}
	\end{equation}
	correspond to the envelopes of the cavity modes, as predicted by the RM. The equation is solved by finite difference discretization (see Appendix). It is apparent that the field envelopes and the eigenfrequencies are close to the corresponding results from a direct solution of the 3D ME. A very important figure to describe the dispersion in multimode resonators is the integrated dispersion\cite{kippenberg2018}, which measures the deviation of the cavity resonances $\omega_m$ from a constant free spectral range (FSR): $D_{int,m} =\omega_m - \omega_0 - \mathcal{D}_1 m$. This quantity is shown in Fig. \ref{fig:nanobeam_cavity}(c). The FSR is fixed to $\mathcal{D}_1=\omega_1-\omega_0$ from the solution of the 3D ME, where $\omega_0$ is always the first eigenvalue (the fundamental mode); this implies that we only consider the relative error on the eigenvalues between the 3D ME results and the RM predictions. Let us also note that the modes are ordered with decreasing frequencies, because the valence band has an upper bound. The deviation $\sigma_m=|\omega_m(3D ME) - \omega_m(RM)|$ is shown in Fig. \ref{fig:nanobeam_cavity}(e): as can be seen, its value is of about 100 GHz.\\ 
	%
	%
	%FIGURE 3

	%
	%
	% calibration
	Let us now allow the polynomial coefficients $\mathcal{C}_i$ to be adjusted, in order to minimize the mismatch between the eigenfrequencies obtained from the solution of the 3D ME and from the RM, namely:
	\begin{equation}
	\epsilon=\dfrac{1}{N} \sum^N_m \left|\dfrac{ \sigma_m }{\omega_0}\right|^2
	\label{eq:err_function}
	\end{equation}
	The results correspond to the red lines in Fig. \ref{fig:nanobeam_cavity}(b), which are now much closer to the Bloch modes obtained from the solution of the 3D ME. This is even more visible when inspecting $D_{int}$, Fig. \ref{fig:nanobeam_cavity}(c), and the corresponding residual, Fig. \ref{fig:nanobeam_cavity}(e), which is now below 10 GHz. Let us now analyze the change of the polynomial coefficients, by inspecting the change of the dependence of $\mathcal{P}$ on $w$ in Fig. \ref{fig:RM_vs_parameter}(b-f).  The relative change of the parameters is very small, and merely appears as an offset. The relative change of $\mathcal{D}_2$ is larger, but this parameter represents a higher-order correction to the coupled wave model. Thus, a slight adjustment of the parameters is sufficient to let our reduced model to converge to the solution of the 3D ME. We will refer hereafter to it as the "updated" RM. The fact that a correction of the parameters is needed is justified by the fact that the adiabatic condition for the tapering is only partially satisfied. Yet, it is noteworthy that the RM already generates a very good approximation of the numerically exact result, and that a slight change of the polynomial coefficients is enough to match the exact result within the discretization error in the solution of the 3D ME. The updated RM solution is in this sense fully equivalent to the 3D ME solution.
	\section{Design of a multimode resonator with a flat dispersion}
	%
	%
	%FIGURE 4
		\begin{figure}[h!]
		\begin{minipage}[c]{0.5\textwidth}
			\includegraphics[width=\textwidth]{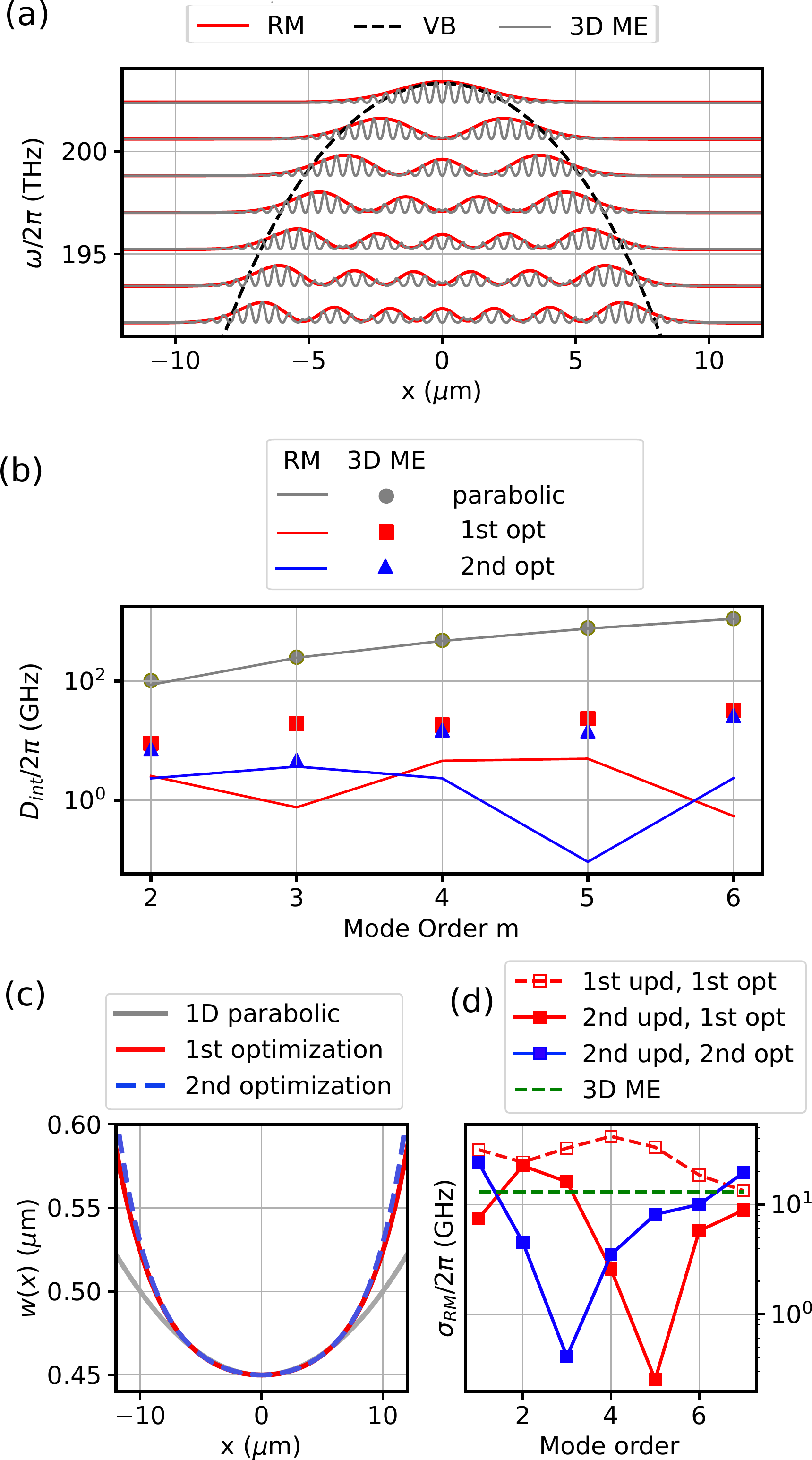}
		\end{minipage}\hfill
		\begin{minipage}[c]{0.48\textwidth}
		\caption{Design of a nanobeam multimode resonator with flat dispersion. (a) Bloch modes (from the 3D ME) and envelopes from the RM after optimisation (as in Fig. \ref{fig:nanobeam_cavity});  (b) Integral dispersion $D_{int}$ for the resonator with parabolic tapering $w_0(x)$ (gray), after the first $w_1(x)$ (red) and the second $w_2(x)$ (blue) optimization, respectively. Markers represent the solution of the 3D ME, and lines correspond to the RM solutions. Note the logarithmic vertical scale. (c) Corresponding tapering profiles $w_i(x)$; (d) residuals (difference between RM and 3D ME) computed after updating the RM with $w_0(x)$ (red dashed), after second update and $w_1(x)$ (red solid) and with $w_2(x)$ (blue). 
}
\label{fig:optimization}
		\end{minipage}
	\end{figure}

\begin{figure}
		\includegraphics[width=0.5\columnwidth]{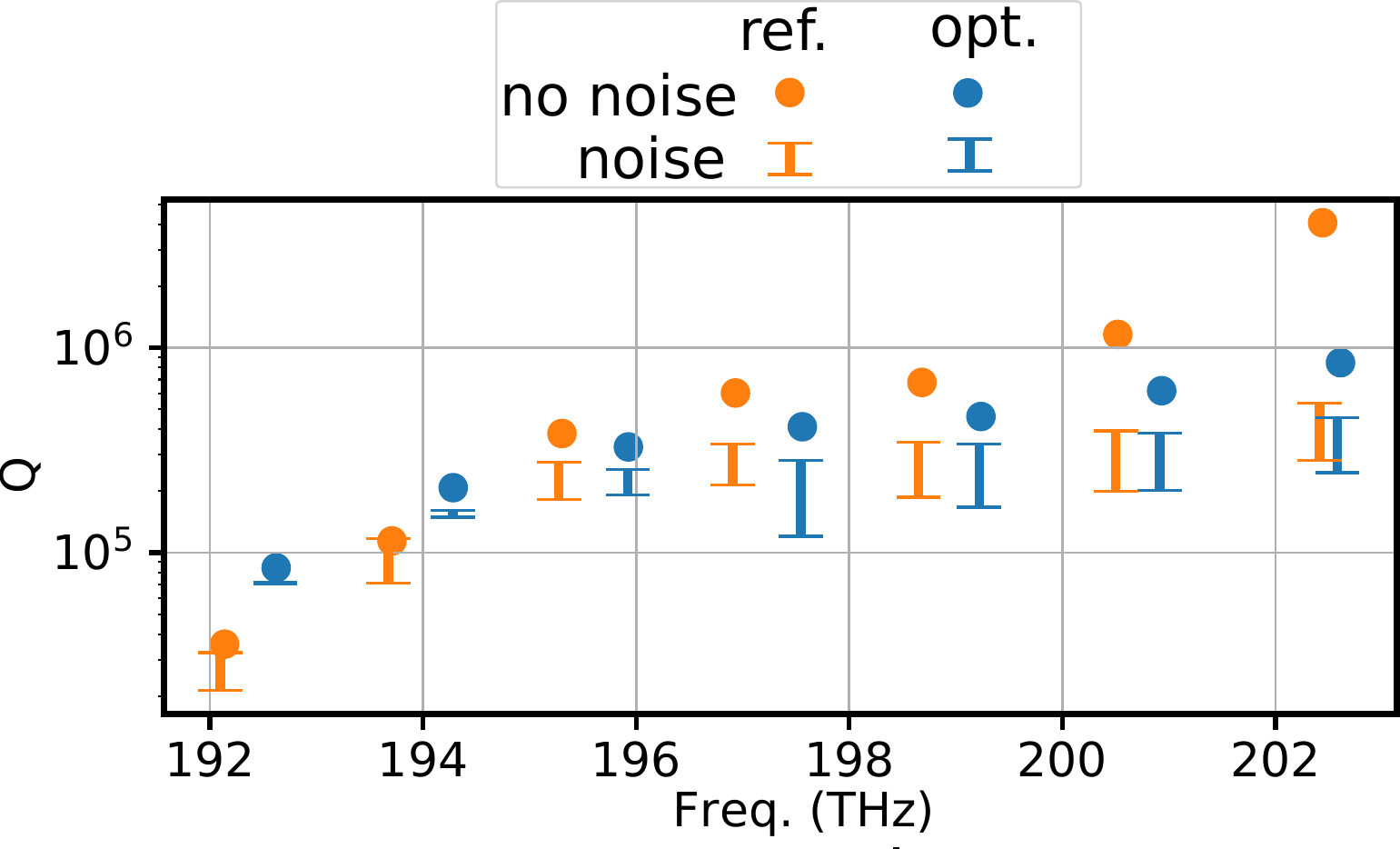}
		\caption{\label{fig:Q_statistics} Calculated Q-factors for the reference cavity with parabolic tapering $w_0(x)$ (orange) and the dispersion-flattened cavity with an optimized profile (blue). Circles correspond to the radiation-limited Q, error bars are an estimate of Q accounting for fabrication disorder. On the x-axis is reported the frequency of the resonances. } 
	\end{figure}

	Let us now consider the updated RM, which consists of the eigenvalue equation \ref{eq:RM_cavity} with the operator \ref{eq:RM_operator}, and parameters defined by the updated coefficients $\mathcal{C}_i$. The profile $w(x)$ is now allowed to change, so that the integrated dispersion $D_{int}$ converges towards a prescribed target. As a notable example, we consider a flat dispersion profile for the first 7 modes as a target, i.e. $D_{int,m}=0$ for $m=1,\dots,7$. The tapering profile is defined by a polynomial with even orders up to $2M=6$: $w(x)=\rho_0 + \sum_{l=1}^M \rho_l x^{2l}$. By this choice we have three degrees of freedom in the optimization process ($\rho_l$, with $l=2,4,6$), ensuring both convergence and high computational efficiency. The cost function $\sum_m |D_{int,m}|$ is minimized with respect to the parameters $\rho_l$. This results into a new profile $w_{1}(x)$, Fig. \ref{fig:optimization}(c),  for which the RM predicts $D_{int}/2\pi$ decreasing by almost 2 orders of magnitude to about 2-4 GHz, Fig. \ref{fig:optimization}(b). The 3D ME are solved again with $w_1(x)$ and the resulting $D_{int}/2 \pi$ is reduced to about 10 GHz, i.e. not as much as the prediction of the RM. The coefficients $\mathcal{C}$ are updated such that RM approaches the 3D ME, as in the previous section. This is necessary since $w_1(x)$ has considerably departed from a parabola. Indeed, panel (d) shows that the residuals (red dashed line) are larger than the estimated accuracy of the 3D ME (green dashed line), yet they decrease below it after the second update (solid red line).
	A new optimized profile is then generated $w_2(x)$, yet no appreciable change is achieved (panels b,c,d), indicating that the procedure has reached convergence, which is essentially set by the accuracy of the 3D ME solver.
	In summary, the method has required one solve of the periodic 3D ME, and two additional 3D ME solves for the cavity, the third one only confirms convergence. \\
	%
	% The details of the numerical implementation and the associated computing time are presented in the Appendix.
	Finally, we analyze how the ID procedure affects the Q factors. Since our procedure does not consider Q as a target for optimisation, there is no guarantee that high-Q values are preserved. This is examined in Fig. \ref{fig:Q_statistics}. The Q factors have been calculated either deterministically (circles) or by modeling the fabrication imperfections by introducing disorder, i.e. by randomly varying (r.m.s  0.5 nm) the position and the diameter of all the holes. The error bars in the figure represent statistics (mean value and the standard deviation of a log-normal distribution) over an ensemble of 20 simulations. Thus, circles correspond to the radiation-limited Q, which decreases Q to less than $10^6$ after the optimisation. Importantly, this is no longer true when disorder is taken into account, as Q factor are basically unchanged.
	% figure here with the statistics of the auqlity factors
	%FIGURE 5
	\section{Conclusions}
	We have introduced a procedure for the inverse design of the dispersion of a multimode nanoscale resonator. The main idea behind our approach harnesses the fact that nanoscale cavities with large quality factors are in general designed according to the principle of  the "gentle confinement". This implies that they can be described as almost adiabatically tapered periodic structures. Inspired by well-known methods of solid state physics, we have introduced a reduced model which is able to capture very well the dispersion of the nanoscale structure in the spectral domain of interest, where the cavity modes exist. The reduced model consists of a linear operator acting on complex functions of a single variable, whose parameters are slowly varying. These parameters are initially determined by fitting the dispersion of a reference periodic structure via a function of the "control" parameter. Here we consider the case of a "nanobeam" photonic crystal cavity. The width of the nanobeam is decreased in the middle, which creates a confining potential in the valence band of the photonic crystal. Next, the model is updated by adapting the nanobeam parameters to the cavity. This two-step procedure avoids issues related with the possible presence of sub-optimal minima in the fitting procedure. We show that the reduced model is equivalent to the solution 3D Maxwell equations within the discretization accuracy of the numerical solver, but it is three orders of magnitude faster. For this reason, any optimisation algorithm can be used. As an example, we consider the problem of flattening the dispersion of a nanoscale resonator. The integrated dispersion $D_{int}$ is reduced below  $\sim$10 GHz, essentially limited by the numerical accuracy of the Maxwell solver.\\
	Our method can be applied to any cavity geometry which can be described via the one or two-dimensional tapering of a periodic structure. The model can be extended to use more than one control parameters, and could also leverage on the presence of multiple waves. In contrast to topological optimisation or inverse design, which are intended to solve a very general class of problems, our procedure is particularly suited to a specific but important class of optical resonators, and could considerably help with the development of nanoscale optical combs, mode-locked lasers, and special purpose parametric generators of non-classical light. 
	%
	%
	%  					APPENDIX
	\section*{Appendix}
	\subsection{Numerical Implementation}
	The reduced model requires the solution of the eigenvalue problem \ref{eq:reduced_model}, which is a system of linear partial differential equations. This is solved by finite-difference discretization of the operator $\hat{\mathcal{L}}_{RM}$ 
	
	\begin{equation}
	\nonumber	
	\hat{\mathcal{L}}_{RM} =
	\begin{bmatrix}
	\hat{\omega}_0-\hat{\mathcal{D}_2}\hat{D}_x^2+\imath\hat{ v}_g\hat{D}_x^{FW} & \hat{\mathcal{K}}\\ \hat{\mathcal{K}} & \hat{\omega}_0-\hat{\mathcal{D}_2}\hat{D}_x^2-\imath \hat{v}_g\hat{D}_x^{BW} 
	\end{bmatrix}\ \ \ \ 
	\label{eq: FD_operator}
	\end{equation}
	\noindent where the hat symbol $\hat{D}_2$, $\hat{D}_x$ means the finite difference approximation of the differential operators, i.e.,  a $2N\times2N$ matrix, where N is the number of points used to approximate the spatial domain. Therefore $\hat{\mathcal{L}}_{RM} $ is  4 $N\times N$ matrix. %For a faster convergence of the method towards interesting phyisical solutions, it is convenient to introduce a frequency offset with respect to the central angular frequency $\omega_0$ in the term $\Delta \omega$ by setting $\Delta \omega=\omega_{ref}-\omega_0$.   % this is a detail and depends on how the algebraic problem is solved
	The operator $\mathcal{L}_{RM}$ is generally non-Hermitian, thus its eigensolutions are not real. Localized eigenfunctions correspond to nearly real eigenvalues ($Re\{\tilde{\omega}\} \gg Im\{\tilde{\omega}\}$).
	The difference operators $\hat{D}^2_x$ and $\hat{D}^{FW,BW}_x$ are implemented on a regular grid $x_j = j\Delta x$ using a second-order central difference scheme, and third-order forward and backward upwind schemes\cite{fdcc,upwind_finite_diff},
	%$\frac{\partial^{(1)}f}{\partial x^{(1)}}\approx\frac{1f(x-2h)-6f(x-1h)+3f(x+0h)+2f(x+1h)}{6h^{1}}$
	%\begin{subequations}
	\begin{align}
		\nonumber
		\hat{D}_x^{FW} f_j &= &\dfrac{-f_{j+2}+6f_{j+1} - 3f_j - 2f_{j-1}}{6\Delta x} \\
		\hat{D}_x^{BW} f_j &= &\dfrac{ f_{j-2} -6f_{j-1} + 3f_j + 2f_{j+1}}{6\Delta x},
		%	&\hat{D}_x^{BW} f(x) &=& \dfrac{f(x-2h)-6f(x-h)+3f(x)+2f(x+h)}{6h}, 
		\label{eq:D_x_upwind}
	\end{align}
	%\end{subequations}
	and, for the second order derivative:
	\begin{equation}
	D_x^2 f_j=\dfrac{f_{j+1} -  2f_j + f_{j-1} }{h^2}\ \ \ .
	\label{eq:D_x_2}
	\end{equation}
	As the terms of the difference scheme outside the domain are implicitly set to zero, these imply Dirichelet boundary conditions, which are not appropriate to represent either evanescent field decay, or dispersive waves. Therefore, the considered computation domain is much larger than the size of the cavity. The other operators are diagonal $\hat{v}_g=\delta_{i,j}v_g(x_j)$, $\hat{\mathcal{\kappa}}=\delta_{i,j}\mathcal{\kappa}(x_j)$, $\hat{\omega}_0=\delta_{i,j}\omega_0(x_j)$, with $\delta_{i,j}$ the Kronecker delta.
	% periodic BC could result from cyclic permutation of the coefficient on the matrix rows 
	%
	%
	All of the code written for optimization and evaluation of the reduced model is written in Julia\cite{bezanson2017}.
	\subsection{Numerical Accuracy}
	A critical issue when calculating the dispersion of multimode nanoscale cavities is that the relative error in the calculation of frequencies can hardly decrease below $10^{-4}$, which translates to inaccuracies of the order of tens of GHz. In ref. \cite{deLasson2018} a variety of methods for solving the ME are compared, for computing the resonances of a nanoscale cavity. It was observed that finite element methods (FEM) converge better than finite differences in time domain (FDTD) methods. Yet, it was concluded that the FEM error is likely to be underestimated, since different implementations of the FEM converge to slightly different results. This underlines how critical the numerical accuracy is with these methods. For this reason, both approaches have been used here. The FDTD algorithm is an in-house code, graphically accelerated with sub-pixel smoothing \cite{Oskooi2009}. The FEM method is implemented within the COMSOL commercial code.\\ 
	% figure here with the convergence of the FDTD
	%FIGURE 6
	\begin{figure}[h!]
		\begin{minipage}[c]{0.5\textwidth}
			\includegraphics[width=\textwidth]{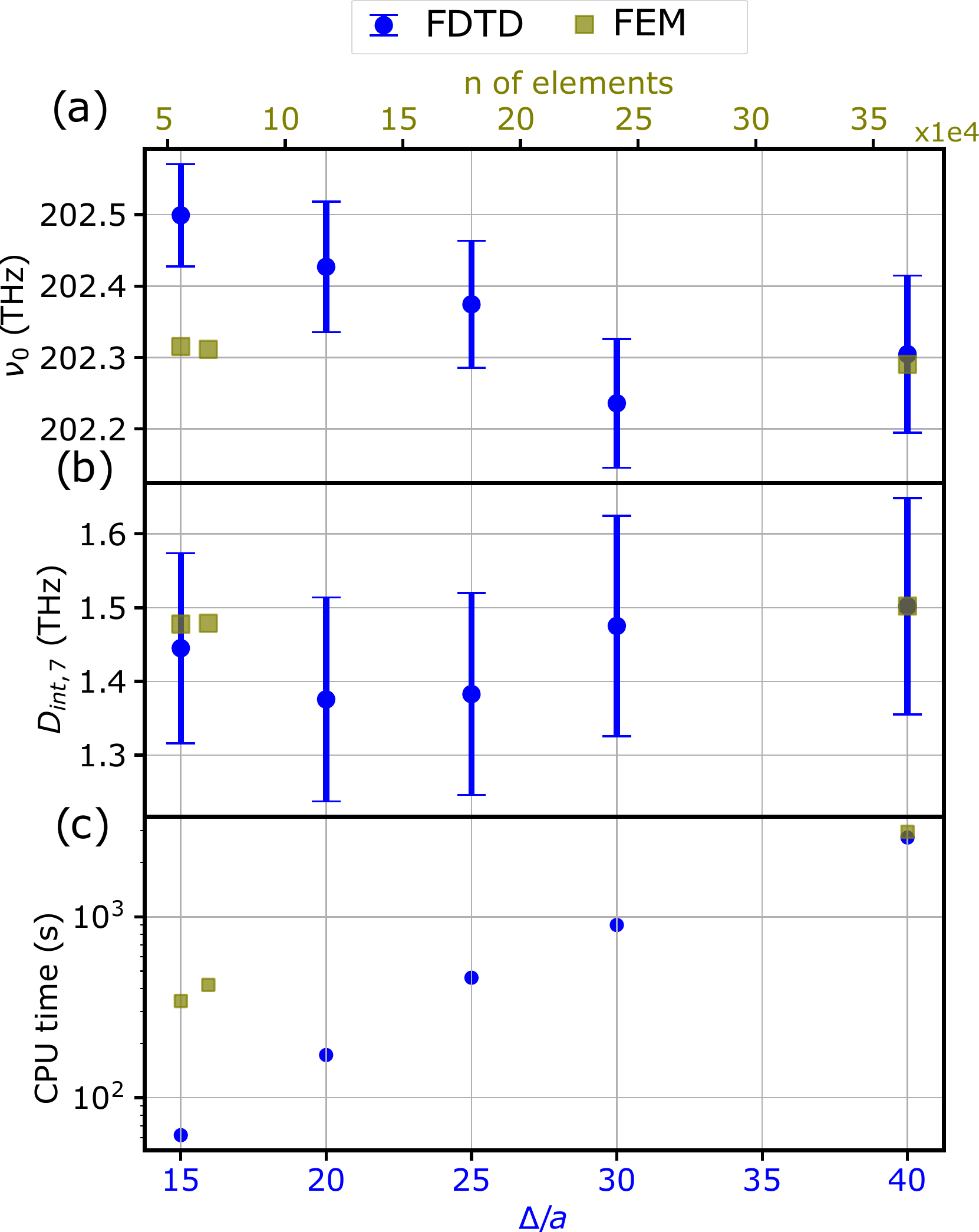}
		\end{minipage}\hfill
		\begin{minipage}[c]{0.48\textwidth}
		\caption{\label{fig:convergence} 						
	Numerical solution of 3D ME with the nanobeam cavity: convergence of FEM (red) and FDTD (blue) methods against number of elements and size of the grid respectively; the error bar stands for the stochastic simulation on an ensemble of 20 realisations. Fundamental mode frequency (a), the integral dispersion up to the $7^{th}$ order mode (b) and the CPU time in units of seconds (c).} 
		\end{minipage}
	\end{figure}

	%
%	The accuracy of the 3D FDTD is estimated to $\sigma_{FDTD}/2\pi=35$ GHz by varying the resolution from $a/30$ to $a/40$. Moreover, disorder, via random fluctuations (with $RMS=1nm$) of the position of the holes, is added and the calculation is repeated over a batch with size between 20 and 60. The frequencies are extracted through a Gaussian fit of the histograms.      	
	Fig. \ref{fig:convergence} compares the two methods by considering the convergence of the frequency of the fundamental order mode $\nu_0$ (a) and the integrated dispersion up to the $7^{th}$ order mode (b) and the time required for the computation (c). Let us note that the computation time scales with $(\Delta x/a)^4$ for the FDTD and moderately superlinear with the number of elements used in the FEM, thus the two horizontal scales cannot be compared directly. Moreover, a more reliable computation of the frequencies through FDTD is obtained by adding random fluctuations in the geometry (as discussed earlier), performing the calculation on 20 to 60 realizations (depending on the resolution) of the structure and considering averages and standard deviations of the histograms of calculated frequencies. The standard deviation depends on the disorder introduced ($rms=1nm$) and it is not an estimate of a numerical accuracy. Thus, the time required by FDTD is much longer indeed, if this method is used. 
	Panel (a) shows that the inaccuracy for the frequency is about 10 GHz when FEM is used. More precisely, the accuracy  assessed by comparing the resonances computed by using either a low  ($67252$ elements) or a high ($364550$ elements) resolution on the tetrahedral mesh, namely $\sigma_{FEM}^2=\dfrac{1}{N}\sum_{m}^{N}|\omega_{FEM,H,m} - \omega_{FEM,L,m}|^2$, hence $\sigma_{FEM}/2\pi = 13$ GHz. As shown in Fig. \ref{fig:nanobeam_cavity}, this is comparable with the average residual between the frequencies computed with our RM and with FEM. The FDTD converges to the same value with $a/\Delta x=40$.  The inaccuracy on the integrated dispersion, panel (b), is similar, and it is matched by FDTD for $a/\Delta x=30$. The two methods give almost identical results for $a/\Delta x=40$. In terms of computing time (for a single FDTD realization) the resolution $a/\Delta x=40$ corresponds to the high resolution FEM mesh, while the low resolution mesh corresponds to $a/\Delta x=25$. With this resolution, we deduce from panels (a,b) that the error of the FDTD is about 100 GHz.\\	
	We conjecture that the accuracy of the RM may be better than that of the numerical solution of 3D ME. As a matter of fact, the discretization of space through finite differences or finite elements results into uncorrelated deviations of the resonances. This source of randomness should vanish in the exact solution, and it may be much reduced in the RM, since the method inherently averages out random deviations. However, the proof of this is problematic, because of the accuracy limitation of numerical solutions. Still, our method guarantees that dispersion flatness remains within the order of 10 GHz in terms of the integrated dispersion, over the first seven confined modes, which is clearly better than what is achievable with the strictly bichromatic design that was reported in \cite{marty2021, combrie2017}, or with the parabolic tapering design of ref \cite{marty2019}. Moreover, here we have shown how, by means of the RM, it is possible to tailor the dispersion of a wide class of resonators, regardless of their initial dispersion curves. Specifically, we could flatten the integrated dispersion of the highest-order modes by almost three orders of magnitude, i.e., from $\ge 1$THz down to $\sim$10 GHz, by drastically changing the $D_{int}$ curve.
	\subsection{The optimization algorithm: workflow and performances}
	\begin{figure}[h!]
		\begin{minipage}[c]{0.5\textwidth}
			\includegraphics[width=\textwidth]{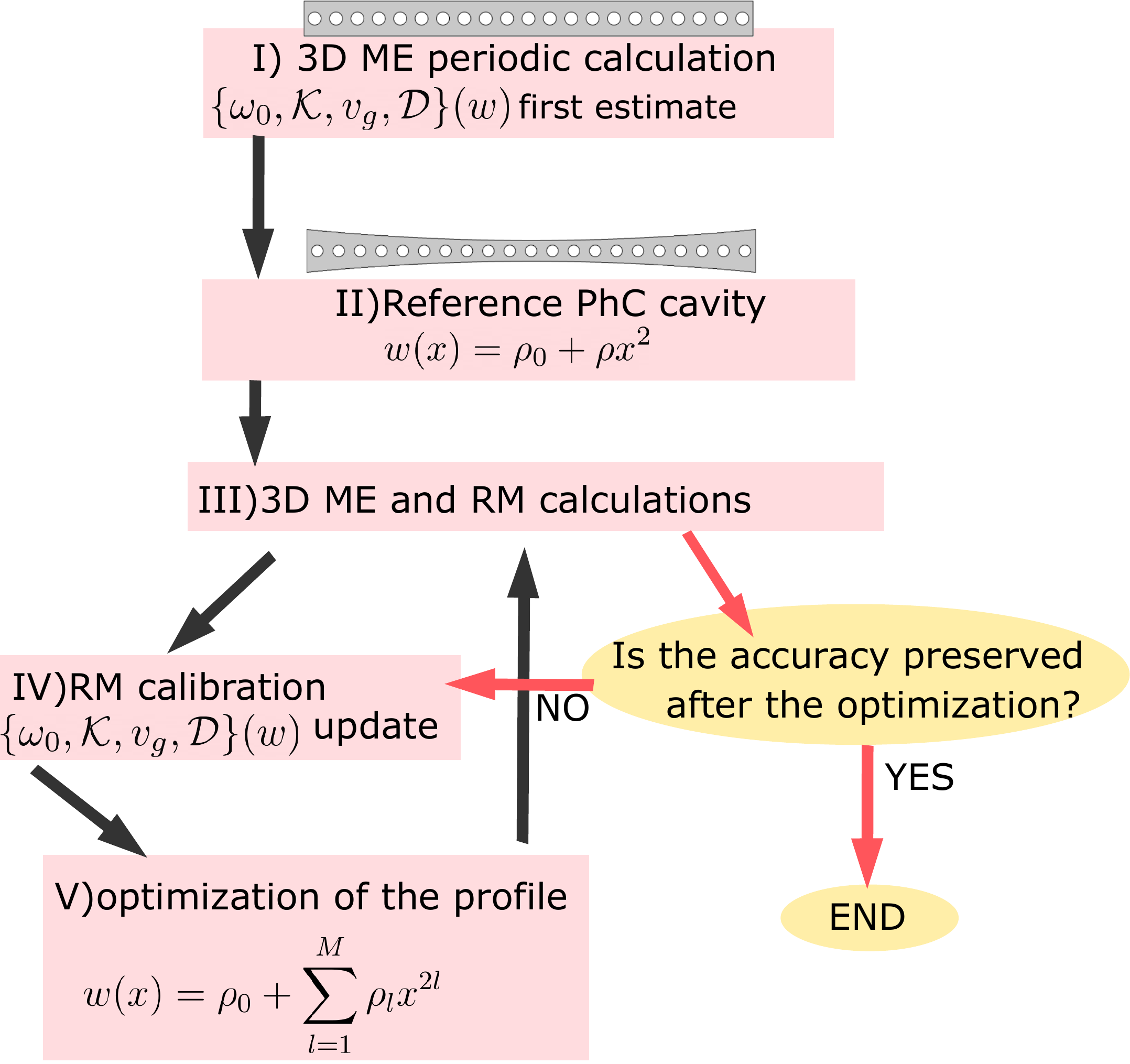}
		\end{minipage}\hfill
		\begin{minipage}[c]{0.48\textwidth}
\caption{\label{fig:flowchart} 						
	Flowchart of the design algorithm: the band diagram calculation of a perfectly periodic structure (I) gives a first estimate of the structure parameters. A reference $w(x)$ PhC profile is taken into account (II) and numerically solved (III). After the calibration of the RM (IV), we proceed with the actual optimization (V). A loop across steps (IV-V) might be necessary, if the accuracy is not sufficient upon optimization (red arrows).} 
		\end{minipage}
	\end{figure}

	\begin{table}%[h!]
		\centering
		\begin{tabular}{|p{27mm}||p{10mm}|p{10mm}|p{10mm}|p{10mm}|p{10mm}|p{10mm}|}
			\hline
			&$w_0$ &$w_0$ upd. & $w_1$ &$w_1$ upd. & $w_2$\\
			\hline\hline
			$\sigma_{RM}$(GHz)& \tikz[baseline]{\node(a11) {200};}& \tikz[baseline]{\node(a12) {8};}& \tikz[baseline]{\node(a13) {30};} & \tikz[baseline]{\node(a14) {12};} & \tikz[baseline]{\node(a15) {13};} \\
			$\Delta^{(2)}$(\%FSR)&\tikz[baseline]{\node(a21) {27};} & \tikz[baseline]{\node(a22) {31};} & \tikz[baseline]{\node(a23) {1.7};}&\tikz[baseline]{\node(a24) {\ 0.17};}&\tikz[baseline]{\node(a25) {0.12};}\\
			\hline
			3D ME &period. &$w_0$ &\multicolumn{2}{c|}{$w_1$} & $w_2$\\
			 CPU time (s)& 80314 &2948&\multicolumn{2}{c|}{2776}&2816\\
			\hline
			fun. calls &\ \ \ \ - & 723 & 220 &557&254\\
			CPU time (s)&\ \ \ \ - & 356 & 107 &261&125\\
			\hline
		\end{tabular}
		\caption{\label{tb:perf}Workflow and overall performances of the method. The reduced model is built from the periodized structure ($w_0$); RM is updated by comparison with the 3D ME calculation (FEM) of the reference cavity ($w_0$ upd.); profile is modified to one with flat dispersion as target ($w_1$); RM is updated again against 3D ME calculation of the optimized cavity ($w_1$ upd.); the geometry is optimized again with the updated model ($w_2$). $\sigma_{RM}$ is the averaged residual between RM and 3D ME, $\Delta^{(2)}=(N-2)^{-1}\sum_j \Delta^{(2)}_j$, with  $\Delta^{(2)}_j=\omega_{j+1}+\omega_{j-1}-2\omega_{j}$ is the averaged second order dispersion, normalized to the FSR. Note that both the update and the optimization steps result in a minimization of two different figures of merit, i.e.  $\sigma_{RM}$ (blue arrows) and  $\Delta^{(2)}$ (red arrows), respectively. The computing time for the 3D FDTD periodic and the FEM calculations for $w_0$, $w_1$ and $w_2$. The time needed to optimize the profile using the RM and the number of calls needed to converge.
		}
		\begin{tikzpicture}[overlay]
		\path[blue, ultra thick,->](a11) edge (a12);
		%\path[thick,->](a12) edge (a22);
		\path[red, ultra thick,->](a22) edge (a23); 
		%\path[thick,->](a23) edge (a13);
		\path[blue, ultra thick,->](a13) edge (a14);
		%\path[thick,->](a14) edge (a24);
		\path[red, ultra thick,->](a24) edge (a25); 
		\end{tikzpicture}
	\end{table}
	The performances of our method are summarized in Table \ref{tb:perf}. Computation is performed using a 32 cores CPU, AMD EPYC 7351, with clock frequency equal to 2.4 GHz, and 64 GB RAM. The first step consists in establishing the RM, which requires the calculation of the bands as a function of the control parameter (here $w$). This takes $\sim3000$ seconds for each of the $26$ values of $w$. The second step is the calculation of the frequencies for the reference cavity, as well as of the first and second optimized geometries, which takes about 3000 s each by using FEM. An error minimization iterative procedure based on the steepest gradient is performed twice for updating the RM, and twice for the optimization of the profile $w(x)$. Convergence requires about few hundreds function calls, and the average time for evaluating the RM is 0.6 s. This gives a sense of the acceleration provided by replacing the solution of the 3D ME with that of the RM. \\ 
	\noindent In FIG.\ref{fig:flowchart} we sketch a flowchart, showing each step of our design technique: the initial step (1) consists in the band diagram calculation of a perfectly periodic structure for different values of the control parameter. A first estimate of the structure parameters and of their dependence on $w$ can be extrapolated by means of a polynomial fit, as sketched in FIG.\ref{fig:RM_vs_parameter}. Consequently, a reference profile of $w=\rho_0 + \rho x^2$ is taken into account (II), and the structure is solved by means of both RM and 3D ME solvers (III). At this point, the RM is calibrated, and the set $\{\omega_0, \mathcal{K}, v_g, \mathcal{D}\}(w)$ updated to recover the best accuracy(IV). The following step is the actual optimization of the $\rho_i |_{i>0}$ coefficients of an even M-th order polynomial expansion $w(x)=\rho_0 + \sum_{l=1}^M \rho_l x^{2l}$ (V). After the optimization cycle, it is essential to verify if the RM accuracy was degraded (red arrows): if so, an other loop of the step (IV-V) will be needed in order to recover the prescribed accuracy.
	\subsection*{Acknowledgments}
	The project leading to this application has received funding from the European Union’s Horizon 2020 research and innovation programme under the Marie Skłodowska-Curie EID project "MOCCA" (G.A. No 814147), the ITN project "OPHELLIA" (G.A. No. 101017136), and the ERC project "STEMS" (G.A. No. 740355). The authors thank Loredana Maria Massaro and Fabrice Raineri for stimulating discussions.

	\bibliography{bibliography,biblio_inversedesign}

\end{document}